\documentclass{article}

\usepackage{arxiv}

\usepackage[utf8]{inputenc} 
\usepackage[T1]{fontenc}    
\usepackage{url}            
\usepackage{booktabs}       
\usepackage{amsfonts}       
\usepackage{nicefrac}       
\usepackage{microtype}      
\usepackage{graphicx}
\usepackage{doi}
\usepackage[all]{xy}
\usepackage{setspace}
\usepackage{xcolor}
\usepackage{tikz}
\usetikzlibrary{cd}

\title{Contact Geometry of Relativistic Particle Motion}

\author{Beg{\" um} Ate\c{s}li\\
Department of Mathematics, Gebze Technical University,\\
41400 Gebze, Kocaeli, Turkey\\
Corresponding author: b.atesli@gtu.edu.tr\\
\And O{\u g}ul Esen\\
Department of Mathematics, Gebze Technical University,\\
41400 Gebze, Kocaeli, Turkey\\
 oesen@gtu.edu.tr\\
 Center for Mathematics and its Applications,  \\ 
Khazar University, Baku, AZ1096, Azerbaijan
\And Michal Pavelka \\
 Mathematical Institute, Faculty of Mathematics and Physics, Charles University,\\ 
 Sokolovsk\'{a} 83, 18675 Prague, Czech Republic\\
  pavelka@karlin.mff.cuni.cz
}

\newcommand{\shorttitle}{Contact-geometric relativistic motion}

\newcommand{\XX}{\mathbb{X}}

\newcommand{\diff}{\mathrm{d}}

\newcommand{\MMM}{\mathcal{M}}

\usepackage{amsmath,amssymb,amsfonts,latexsym,float,graphics,epsfig}

\begin{document}
\maketitle

\begin{abstract}
We introduce a new geometric framework for relativistic particle dynamics based on contact geometry and suitable for treating dissipative processes like particle decay. The dynamics is formulated on a nine--dimensional extended phase space consisting of four position coordinates, four momenta, and an additional variable (functioning as a geometric variant of the particle's proper
time). In this setting, the evolution is generated by an evolution contact vector field with a contact Hamiltonian encoding the mass shell. 
By promoting the proper time to an independent variable, the relativistic Hamilton canonical equations are rewritten in a fully geometric form without having to identify the proper time with a parameter along the worldlines. This makes for instance the evolution of massless particles (photons) well-defined without the need of reparametrization. 
The framework is then applied to decaying particles. 
Finally, we formulate a covariant kinetic theory and show how decaying particles can be described geometrically in this framework, changing the entropy.
\end{abstract}



\section{Introduction}

Dynamics of a relativistic particle can be written as Hamilton canonical equations where the evolution is parametrized by the proper time of the particle \cite{hartle}. However, this picture is not compatible with the usual geometric interpretation of Hamiltonian mechanics, giving the Hamiltonian vector field and its integral curves (worldlines), but keeping the parametrization of the motion along the worldlines arbitrary. In this work, we introduce a new geometric framework for relativistic particle dynamics based on contact geometry, which allows us to write the evolution equations in a fully geometric form without having to identify the proper time with a parameter along the worldlines. This makes for instance the evolution of massless particles (photons) well-defined without the need of reparametrization. Moreover, the framework sheds light on the entropy evolution in relativistic kinetic theory.

Throughout the manuscript, we shall combine three types of geometries: symplectic geometry, contact geometry, and Riemannian geometry. There exists a symbiotic relationship between physical systems and the geometric structures in which they are realized. Hamiltonian dynamics attains its elegant formulation through symplectic and Poisson geometry, see, for example, \cite{abraham-marsden,arnoldbook,LiMa}. A similar interplay arises between contact geometry and thermodynamics, where irreversible and energy-dissipative processes find a natural geometric description \cite{Bravetti19,esen2021implicit,EsGrPa22a, mg2014,Mrugala,Rajeev,simoes2020contact}.
A geometric formulation of relativistic particle dynamics was also given in \cite{tulczyjew-relativity}, but in the present manuscript we provide a clearer way within contact geometry.
Finally, Riemannian geometry is a natural setting for relativistic physics \cite{einstein-gr,Carroll,MTW-73,Wald}, and the here-presented contact formulation will reduce to it.

The connection between geometric mechanics, thermodynamics, and relativistic motion may provide more insight into unresolved problems of general relativity, such as the existence of dark matter and energy \cite{debono2016,van-abe,pszota2024}, or to numerical modeling of general relativity \cite{dumbser2025-einstein}. 

In this manuscript, we take the evolution contact vector field generated by a contact Hamiltonian encoding the mass shell. The coordinate along the Reeb vector field (part of the evolution contact vector field) will be seen as a variant of the particle's proper time. Therefore, the evolution takes place on a nine-dimensional manifold of four-position, four-momentum, and the proper time. The dynamics is fully geometric and does not require any identification of the proper time with a parameter along the worldlines.
Then, we formulate a contact kinetic theory that is suitable for the description of decaying particles in a fully geometric way. 



Section \ref{sec.geo} recalls the basics of contact geometry and the evolution contact vector field. In Section \ref{sec.rel}, we apply this framework to relativistic particle dynamics. Section \ref{sec.kin} contains a covariant kinetic theory generated by the evolution contact vector field, including the entropy behavior. Finally, in Section \ref{sec.app}, we show how the framework works in the case of special relativity, Newtonian gravity, and decaying particles.

\section{Hamiltonian Dynamics on Contact Manifolds}\label{sec.geo}

Let us first recall contact geometry and the evolution contact vector field \cite{arnoldbook,LiMa87}. 
A $(2n+1)$-dimensional manifold $M$ admitting a one-form $\eta$ satisfying the non-degeneracy condition
\begin{equation}
    d\eta^{\,n} \wedge \eta \neq 0 
\end{equation}
is called a contact manifold.
On a contact manifold, there exists a Reeb vector field $\mathcal{R}$ encoding the non-integrability, determined by the conditions
\begin{equation}
    \iota_{\mathcal{R}}\eta = 1,
    \qquad
    \iota_{\mathcal{R}} d\eta = 0 .
\end{equation}
For a Hamiltonian function $H$ on $T^*Q\times \mathbb{R}$, there is a corresponding evolution contact vector field $\XX_H$, defined as follows \cite{Simoes-thermo}
\begin{equation}
\iota_{\XX_{H}}\eta =0,\qquad \iota_{\XX_{H}}d\eta =dH-\mathcal{R}(H) \eta   \label{contact}
\end{equation} 
with the following properties:
\begin{equation}\label{L-X-eta}
    \mathcal{L}_{\XX_H}\eta
    = dH-\mathcal{R}(H)\eta,
    \qquad
    \mathcal{L}_{\XX_H}\big(d\eta^{\,n}\wedge\eta\big)
    = - n\, \mathcal{R}(H)\, d\eta^{\,n}\wedge\eta,
    \qquad
    \mathcal{L}_{\XX_H} H
    = 0.
\end{equation}
Thus, the dynamics preserves neither the contact one-form (even up to a conformal factor) nor the volume form \(d\eta^{\,n}\wedge\eta\) in general, 
as the divergence of the evolution contact vector field is
\begin{equation}\label{div-X-H}
    \mathrm{div}(\XX_H) = - n\, \mathcal{R}(H).
\end{equation}
The last identity in \eqref{L-X-eta} shows, however, that the Hamiltonian is constant along the evolutionary trajectories. 

Another possibility is to use the contact canonical vector field which preserves the contact one-form, but this vector field does not preserve the Hamiltonian unless $H=0$ \cite{lie-geometrie,arnoldbook,bravetti}. Here, however, we prefer evolution contact vector field, since it preserves the Hamiltonian and is often more suitable for thermodynamics \cite{omg1}.

A generic example of a contact manifold is obtained by the contactization of the canonical symplectic manifold. In this paper we realize this construction as follows. Let $Q$ be the $4D$ spacetime manifold with coordinates $(q^\mu)$, where the index $\mu$ runs from $1$ to $4$, and the extended cotangent bundle $T^*Q\times \mathbb{R}$ equipped with the Darboux coordinates $(q^\mu,p_\mu,\phi)$, then we define the contact one-form as
\begin{equation}\label{eta-Q}
    \eta = d\phi - \theta_Q = d\phi - p_\mu\, dq^\mu
\end{equation}
The corresponding Reeb vector field is computed to be $\mathcal{R} = \partial/\partial\phi$. In this picture, 
the evolution contact vector field is computed to be
\begin{equation}\label{con-dyn}
\XX_H=\frac{\partial H}{\partial p_{\mu}}\frac{\partial}{\partial q^{\mu}}  - \left(\frac{\partial H}{\partial q^{\mu}} + \frac{\partial H}{\partial \phi} p_{\mu} \right)
\frac{\partial}{\partial p_{\mu}} + p_{\mu}\frac{\partial H}{\partial p_{\mu}}\frac{\partial}{\partial \phi}.
\end{equation}
Thus, we obtain the evolution contact equations for $H$ as
\begin{equation}\label{conham}
\frac{dq^{\mu}}{d \lambda} = \frac{\partial H}{\partial p_{\mu}}, 
\qquad
 \frac{dp_{\mu}}{d\lambda}  = -\frac{\partial H}{\partial q^{\mu}}- 
p_{\mu}\frac{\partial H}{\partial \phi}, 
\qquad \frac{d\phi}{d\lambda} = p_{\mu}\frac{\partial H}{\partial p_{\mu}},
\end{equation}
where $\lambda$ is an arbitrary parameter along the flow with no geometrical meaning or physical significance.
The third equation (the Hamiltonian action) provides the justification for referring to $\phi$ as the action variable.
Note that in Galilean physics the meaningless parameter $\lambda$ is interpreted as the global time, but we wish to avoid any such identification so that all physical quantities are formulated geometrically.


\section{Relativistic Particle Motion in Contact Framework}\label{sec.rel}
In the standard formulation of relativistic particle mechanics, the physical constraint imposed on the four-momentum is
\begin{equation}\label{eq.constraint.standard}
    g^{\mu\nu} p_\mu p_\nu = - m^{2} c^{2},
\end{equation}
where $g^{\mu\nu}$ is the inverse spacetime metric tensor (with signature $(-,+,+,+)$), 
$m$ is the rest mass of the particle, and $c$ is the speed of light. We consider the usual index-raising and index-lowering identity $    g^{\mu\nu} g_{\nu\alpha} = \delta^{\mu}_{\alpha}
$, where $g_{\mu\nu}$ is the metric tensor. 

In our contact-geometric formulation with a mass $m(\phi)$, that may depend on the variable $\phi$,
the mass-shell condition $H=0$ imposes the constraint
\begin{equation}\label{eq.constraint}
    g^{\mu\nu} p_\mu p_\nu = - m(\phi)^{2} c^{2}.
\end{equation}
This expresses the fact that the four-momentum $p_\mu$ lies on a generalized
mass--shell determined by both the metric and the $\phi$-dependent mass.
When $m(\phi)=m$ is constant, this reduces to the standard constraint \eqref{eq.constraint.standard}.
The subset of the extended phase space where this constraint is satisfied will be denoted by
\begin{equation}\label{MMM}
    \MMM = \left\{ (q,p,\phi) \,\middle|\, g^{\mu\nu}(q) p_\mu p_\nu = - m(\phi)^{2} c^{2} \right\}.
\end{equation}
We shall respect this mass-shell constraint in this manuscript, but note that it is possible to have also states outside of the shell in quantum field theory \cite{peskin}.

Throughout this work, the metric $g_{\mu\nu}(q,\phi)$ is regarded as a fixed, static background field.  
In particular, we do not impose or consider the Einstein field equations governing its dynamical evolution \cite{einstein-gr,landau2}; 
the geometry of spacetime is taken as given, and only the dynamics of the particle is studied.

We choose the following contact Hamiltonian function
\begin{equation}\label{eq.Ham}
    H(q,p,\phi) = \frac{1}{2}\Bigl(g^{\mu\nu}(q,\phi)p_\mu p_\nu + m(\phi)^{2} c^{2}\Bigr),
\end{equation}
where the metric depends on both the position (as usual) and the action variable $\phi$, and $m(\phi)$ is the mass that may depend on the action variable.  
For this contact Hamiltonian, the evolution contact equations \eqref{conham} are computed to be
\begin{subequations}\label{eq.con.can.H}
\begin{eqnarray}
\frac{\diff q^\mu}{\diff \lambda} &=& 
    g^{\mu\nu} p_\nu, \\[0.3em]
\frac{\diff p_\mu}{\diff \lambda} &=&
    -\frac{1}{2}\frac{\partial g^{\alpha\beta}}{\partial q^\mu} p_\alpha p_\beta
    - p_\mu\!\left(\frac{1}{2}\frac{\partial g^{\alpha\beta}}{\partial \phi} p_\alpha p_\beta
            + m(\phi) c^{2} m'(\phi)\right),
    \\[0.3em]
\frac{\diff \phi}{\diff \lambda} &=&
    g^{\mu\nu} p_\mu p_\nu.
\end{eqnarray}
\end{subequations}
Recall that the Hamiltonian is constant along the motion that is $\diff H / \diff \lambda =0$. Therefore, if the motion starts on the mass--shell \(\MMM\), that is, if \(H=0\) initially, then this condition remains valid for all \(\lambda\). Hence, the evolution contact dynamics preserves the constraint surface \(\MMM\): trajectories starting on the shell stay on the shell.
 
\noindent \textbf{Arbitrariness of the Parameter $\lambda$.}
The parameter $\lambda$ in \eqref{eq.con.can.H} may be regarded as an arbitrary parametrization of curves on the contact manifold. Therefore, it is desirable to eliminate this arbitrariness and obtain a reduced system of equations whose evolution does not depend on the particular choice of $\lambda$. Since the physically relevant quantities are the four-position $q^{\mu}$ and the four-momentum $p_{\mu}$, we seek a reformulation of \eqref{eq.con.can.H} in which the derivatives are taken with respect to the remaining scalar variable $\phi$, thereby removing the freedom in the parametrization. 
For this purpose, we provide the following commutative diagram to make the geometrical foundations of our discussion more explicit
\begin{equation}\label{diag}
\xymatrix{
    T(T^*Q\times \mathbb{R})
        \ar[rrrr]^{T\pi} &&&& TT^*Q \\
    T^*Q\times \mathbb{R}
        \ar[drr]
        \ar[u]^{\XX_H}
        \ar[rrrr]^{\pi=\mathrm{pr}_{T^*Q}}
        \ar[rrd]^{\mathrm{pr}_{\mathbb{R}}}
        &&&&
        T^*Q
        \ar[u]^{Y}
        \\
    && \mathbb{R}_\phi
        \ar[urr]^{\bar{\gamma}=\bar{\gamma}(\phi)}
        \\
    &&
        \mathbb{R}_\lambda
        \ar[uull]^{\gamma=\gamma(\lambda)}
        \ar[u]^{\phi(\lambda)}
        \ar[uurr]_{\tilde{\gamma}=\tilde{\gamma}(\lambda)}
}
\end{equation}
In this diagram, we have introduced curves
\begin{equation}
\begin{split}
    \gamma &: \mathbb{R}_\lambda \longrightarrow T^*Q\times\mathbb{R}, \qquad \lambda \mapsto \gamma(\lambda)=(\tilde{\gamma}(\lambda),\phi(\lambda)),
    \\
    \tilde{\gamma}&: \mathbb{R}_\lambda \longrightarrow T^*Q, \qquad  \lambda \mapsto  \tilde{\gamma}(\lambda)    .
\end{split}
\end{equation}

Our goal is to project the evolution contact vector field $\XX_H$ on $T^*Q\times\mathbb{R}$ onto a vector field $Y$ on $T^*Q$ via
\begin{equation}
    Y\circ \pi (z,\phi)
    = T\pi \circ \XX_H(z,\phi).
\end{equation}
Note that the vector field $Y$ is well defined only if $\XX_H$ is projectable. 
If the curve $\gamma(\lambda)$ is an integral curve (flow) of the evolution contact vector field, then it satisfies
\begin{equation}
    \frac{\diff \gamma}{\diff \lambda}
        = \XX_H \circ \gamma(\lambda).
\end{equation}

Referring to the commutative diagram \eqref{diag}, we can argue that for flows in the symplectic manifold $T^*Q$, we can determine the independent variable in two ways depending on $\lambda$ and depending on $\phi$. These are related by 
\begin{equation}
\tilde{\gamma} (\lambda) =\bar{\gamma} \circ \phi (\lambda). 
\end{equation}
Respectively, we have two realizations of the dynamics generated by the vector field $Y$ as
\begin{equation} \label{dyn-Y}
\frac{\diff \tilde{\gamma}}{\diff \lambda} = Y \circ \tilde{\gamma} (\lambda), \qquad 
\frac{\diff \bar{\gamma}}{\diff\phi} = Y \circ \bar{\gamma} (\phi).
\end{equation}

\textbf{Case I: Projectable Vector Fields.}  
The vector field $\XX_H$ is projectable with respect to $\pi$ if its components along the directions 
$\partial / \partial q^\mu$ and $\partial / \partial p_\mu$ depend only on $(q,p)$ and not on the fiber coordinate~$\phi$.  
That is, if $\XX_H$ can be written in the form
\begin{equation}
    \XX_H
    = X^\mu(q,p) \frac{\partial}{\partial q^\mu}
    + X_\mu(q,p) \frac{\partial}{\partial p_\mu}
    + X(q,p,\phi)\, \frac{\partial}{\partial \phi},
\end{equation}
then the first two sets of coefficients define a well-defined vector field on $T^*Q$:
\begin{equation}
    Y
    = X^\mu(q,p) \frac{\partial}{\partial q^\mu}
    + X_\mu(q,p) \frac{\partial}{\partial p_\mu}.
\end{equation}
In this case, the dynamics becomes
\begin{equation}
    \frac{\diff q^\mu}{\diff \lambda} = X^\mu(q,p),
    \qquad
    \frac{\diff p_\mu}{\diff \lambda} = X_\mu(q,p),
\end{equation}
and the projection simply removes the $\phi$–component without altering the evolution of $(q,p)$.

\textbf{Case II: Non-projectable Vector Fields.}  
When the coefficients of $\XX_H$ do depend on $\phi$, projectability fails.  
In this situation, the dynamics of $(q,p)$ cannot be directly projected, but one may remove the arbitrariness of the parameter~$\lambda$ by using the scalar variable~$\phi$ as a new evolution parameter.
To achieve this, observe that $\phi$ is transverse to $\XX_H$ on $M$: the last
equation in \eqref{eq.con.can.H} gives
\begin{equation}
    \diff\phi(\XX_H) = X \neq 0,
\end{equation}
so no integral curve of $\XX_H$ is tangent to a level set $\{\phi = \mathrm{const}\}$.
This transversality is an intrinsic property of $\XX_H$ and $\phi$ on $T^*Q\times \mathbb{R}$—it requires
no auxiliary map or separate invertibility assumption.

Transversality allows one to reparametrize all orbits by $\phi$ canonically.
Geometrically, this amounts to replacing $\XX_H$ by its $\phi$-normalized rescaling
\begin{equation}
    \widetilde{\XX}_H
    = \frac{\XX_H}{\diff\phi(\XX_H)}
    = \frac{X^\mu}{X} \frac{\partial}{\partial q^\mu}
    + \frac{X_\mu}{X} \frac{\partial}{\partial p_\mu}
    + \frac{\partial}{\partial \phi},
\end{equation}
the rescaling of $\XX_H$ satisfying $\diff\phi(\widetilde{\XX}_H) = 1$ with the
same unparametrized orbits as $\XX_H$.  The reduced vector field on $T^*Q$ is then
the pushforward under the bundle projection $\pi$:
\begin{equation}
    Y
    = \pi_*\widetilde{\XX}_H
    = \frac{X^\mu}{X} \frac{\partial}{\partial q^\mu}
    + \frac{X_\mu}{X} \frac{\partial}{\partial p_\mu},
\end{equation}
where $X = \diff\phi(\XX_H)$ is the $\phi$-component of $\XX_H$.
Thus the dynamical equations with $\phi$ as evolution parameter are
\begin{equation}
    \frac{\diff q^\mu}{\diff \phi} = \frac{X^\mu}{X},
    \qquad
    \frac{\diff p_\mu}{\diff \phi} = \frac{X_\mu}{X}.
\end{equation}

\textbf{Application to relativity.}  
For the relativistic contact Hamiltonian system \eqref{eq.con.can.H}, dividing the equations for $q^\mu$ and $p_\mu$ by the equation for $\phi$ and evaluating on the mass shell $H=0$ (where $-g^{\alpha\beta}p_\alpha p_\beta = m(\phi)^{2}\,c^{2}$) gives the reduced evolution equations
\begin{subequations}\label{eq.evo.qp.phi}
\begin{eqnarray}
    \label{eq.evo.q.phi}
    \frac{\diff q^\mu}{\diff \phi}
    &=& \frac{-g^{\mu\nu} p_\nu}{m(\phi)^2 c^2}, \\[0.4em]
    \frac{\diff p_\mu}{\diff \phi}
    &=& \frac{1}{2 m(\phi)^2 c^2}
        \frac{\partial g^{\alpha\beta}}{\partial q^\mu} p_\alpha p_\beta
        + \frac{p_\mu}{2 m(\phi)^2 c^2}
        \frac{\partial g^{\alpha\beta}}{\partial \phi} p_\alpha p_\beta
        + \frac{p_\mu m'(\phi)}{m(\phi)}.
\end{eqnarray}
\end{subequations}

\noindent \textbf{Proper Time vs.\ $\phi$.}
To understand the physical meaning of the additional contact coordinate $\phi$,
let us recall that a relativistic particle carries only one intrinsic notion of
time: its  proper time  $\tau$.  This quantity measures the interval
registered by a clock moving with the particle and is therefore the only
physically meaningful time parameter available.  It is defined by
\begin{equation}\label{eq.tau.def}
    c^2 \diff \tau^2 = -g_{\mu\nu} \diff q^\mu \diff q^\nu .
\end{equation}

A comparison between $\tau$ and $\phi$ can be made using the reduced evolution
equation \eqref{eq.evo.q.phi}.  This equation states that the change of the
particle’s position with respect to $\phi$ is proportional to its momentum.
Substituting this into \eqref{eq.tau.def}, we obtain
\begin{equation}
    -g_{\mu\nu}\diff q^\mu \diff q^\nu
        = \frac{(\diff \phi)^2}{m(\phi)^4 c^4}\, (-g_{\mu\nu} g^{\mu\alpha} g^{\nu\beta} p_\alpha p_\beta)
        = \frac{-g^{\alpha\beta} p_\alpha p_\beta}{m(\phi)^4 c^4}\, (\diff \phi)^2.
\end{equation}
On the mass shell where $H=0$, we have $-g^{\alpha\beta}p_\alpha p_\beta = m(\phi)^2 c^2$, so
\begin{equation}
    -g_{\mu\nu}\diff q^\mu \diff q^\nu = \frac{(\diff \phi)^2}{m(\phi)^2 c^2}.
\end{equation}
The left hand side is positive for a time-like path.

Comparing with \eqref{eq.tau.def}, we fix the orientation by choosing the branch
compatible with standard general-relativistic sign conventions, and obtain
\begin{equation}\label{eq.tau.phi}
    \diff \phi = -m(\phi) c^2\,\diff \tau,
    \qquad
    \frac{\diff}{\diff \phi}
        = -\frac{1}{m(\phi) c^2}\frac{\diff}{\diff \tau}.
\end{equation}
Therefore, the contact coordinate $\phi$ plays the role of a generalized
proper time, with the rate of change governed by the particle's instantaneous rest mass energy.
When $m(\phi)=m$ is constant, this reduces to the relation
$\phi = -mc^2 \tau$ (up to an integration constant).  The additional dimension introduced by contact geometry is thus
not an artificial mathematical device—it directly corresponds to the intrinsic
clock carried by the particle.

Substituting \eqref{eq.tau.phi} into the reduced equations
\eqref{eq.evo.qp.phi}, and using the chain-rule identities
\begin{equation}
    m'(\phi)=-\frac{\dot{m}(\tau)}{m(\tau)c^{2}},
    \qquad
    \frac{\partial g^{\alpha\beta}}{\partial\phi}\bigg|_q
    =-\frac{1}{m(\tau)c^{2}}\frac{\partial g^{\alpha\beta}}{\partial\tau}\bigg|_q,
\end{equation}
we obtain the evolution equations fully in proper time:
\begin{subequations}\label{eq.evo.pq}
\begin{eqnarray}
    \label{eq.evo.q}
    \frac{\diff q^\mu}{\diff \tau}
        &=& \frac{g^{\mu\nu}(q,\tau)\, p_\nu}{m(\tau)},
    \\[0.4em]
    \label{eq.evo.p}
    \frac{\diff p_\mu}{\diff \tau}
        &=& -\frac{1}{2 m(\tau)}\frac{\partial g^{\alpha\beta}}{\partial q^\mu} p_\alpha p_\beta
            +\frac{p_\mu}{2 m(\tau)^2 c^{2}}\frac{\partial g^{\alpha\beta}}{\partial\tau} p_\alpha p_\beta
            +\frac{\dot{m}(\tau)}{m(\tau)}\,p_\mu,
\end{eqnarray}
\end{subequations}
where $\dot{m}\equiv\diff m/\diff\tau$ and $g^{\alpha\beta}=g^{\alpha\beta}(q,\tau)$ is the metric viewed as a field on spacetime parametrised by proper time (the $\phi$-dependence of the metric is re-expressed via $\phi=\phi(\tau)$).

Equation \eqref{eq.evo.q} gives the identification between momentum and four-velocity:
\begin{equation}\label{eq.pu}
    g^{\mu\nu}(q,\tau)\, p_\nu = m(\tau)\, u^\mu,
    \qquad
    u^\mu = \frac{\diff q^\mu}{\diff \tau},
\end{equation}
or equivalently $p_\mu = m(\tau)\, g_{\mu\nu}(q,\tau)\, u^\nu$.  
The $\tau$-dependence of the mass modifies the effective inertial mass of the particle
as measured along its worldline.

The equations \eqref{eq.evo.pq} describe the evolution of the particle in curved spacetime under the influence of both the metric $g_{\mu\nu}$ and the mass $m(\tau)$.  
When $m(\tau)=m$ is constant and the metric has no $\tau$-dependence, the last two terms in \eqref{eq.evo.p} vanish and \eqref{eq.evo.pq} reduces to the standard geodesic equations of general relativity, $\ddot{q}^\mu + \Gamma^\mu_{\alpha\beta}\dot{q}^\alpha\dot{q}^\beta = 0$, where $\Gamma^\mu_{\alpha\beta}$ are the Christoffel symbols \cite{landau2}.  

\section{Kinetic theory}\label{sec.kin}
\paragraph{General formulation.}
At the geometric level, kinetic theory is formulated on a phase-space manifold $M$
with a chosen volume form $\Omega$ and a dynamical vector field $X$.
For any observable $A\in C^{\infty}(M)$, the evolution law is
\begin{equation}
    \frac{\partial A}{\partial \lambda} = \mathcal{L}_{X}A = X(A).
\end{equation}
For a density form $\mu=f\,\Omega$, the dual evolution is
\begin{equation}
    \frac{\partial \mu}{\partial \lambda} = -\mathcal{L}_{X}\mu,
\end{equation}
which, for top-degree forms, is equivalent to the continuity equation
\begin{equation}
    \frac{\partial f}{\partial \lambda} + \mathrm{div}_{\Omega}(fX) = 0,
\end{equation}
equivalently,
\begin{equation}
    \frac{\partial f}{\partial \lambda} + X(f) + f\,\mathrm{div}_{\Omega}(X) = 0.
\end{equation}
This is the general Liouville equation in divergence form, independent of the
particular geometric structure.

Integrating the continuity equation gives (by the Cartan magic formula and the Stokes theorem \cite{fecko})
\begin{equation}
    \frac{\diff}{\diff \lambda}\int_{M} f\,\Omega
     = -\int_{\partial M}\iota_{X}(f\,\Omega).
\end{equation}
Hence total probability $\int_{M}f\,\Omega$ is conserved on closed phase spaces,
or more generally under vanishing boundary flux.

\paragraph{Evolutionary contact Liouville equation.}
In our setting, $M=T^*Q\times\mathbb{R}$ and the dynamics is generated by the
evolution contact vector field $X=\XX_H$, with the canonical contact volume form
$\Omega=\eta\wedge (\diff\eta)^4$.  The continuity equation therefore becomes
\begin{equation}
    \frac{\partial f}{\partial \lambda} + \mathrm{div}(f \XX_H) = 0,
\end{equation}
which can be expanded as
\begin{equation}\label{eq.Liouville}
    \frac{\partial f}{\partial \lambda} + \XX_H(f) + f \mathrm{div}(\XX_H) = 0.
\end{equation}
Using the divergence of the evolution contact vector field, see Eq.~\eqref{div-X-H} with $n=4$, we arrive at
\begin{equation}
    \frac{\partial f}{\partial \lambda} + \XX_H(f) - 4 f \frac{\partial H}{\partial \phi} = 0.
\end{equation}
This is the contact Liouville equation, which describes the evolution of the distribution function $f$ on the extended phase space $T^*Q \times \mathbb{R}$. In coordinates, this can be written explicitly as
\begin{equation}
    \frac{\partial f}{\partial \lambda} 
   + \frac{\partial H}{\partial p_{\mu}}\frac{\partial f}{\partial q^{\mu}} 
   - \left(\frac{\partial H}{\partial q^{\mu}} + \frac{\partial H}{\partial \phi} p_{\mu}\right)\frac{\partial f}{\partial p_{\mu}} 
   + p_{\mu}\frac{\partial H}{\partial p_{\mu}}\frac{\partial f}{\partial \phi} 
   - 4 f \frac{\partial H}{\partial \phi} = 0.
\end{equation}
In relativistic kinetic theory, $f$ is usually taken as a scalar field on phase space and therefore has no explicit dependence on the trajectory parameter $\lambda$. In that autonomous case, $\partial f/\partial \lambda=0$ and the kinetic equation is written intrinsically as
\begin{equation}
    \XX_H(f) - 4 f \frac{\partial H}{\partial \phi} = 0.
\end{equation}
If we assume, moreover, that $H$ and $f$ are independent of $\lambda$ and $\phi$, we obtain the usual relativistic kinetic equation in \cite{synge,vereshchagin-aksernov}.

On the mass shell $\MMM$ defined in Eq. \eqref{MMM}, where $H=0$, the term 
\begin{equation}
   \frac{\partial H}{\partial \phi} \Big|_{H=0} = \frac{1}{2}\frac{\partial g^{\alpha\beta}}{\partial \phi} p_\alpha p_\beta + m(\phi)c^2 m'(\phi)
\end{equation}
does not specifically vanish. This means that the evolution equation for the distribution function $f$ itself is not in the conservative form; the evolution for the whole top-form $f\,\Omega$ is conservative. 

\paragraph{Entropy production and the $H$-theorem.}
Consider now the total entropy 
\begin{equation}\label{eq.entropy}
   S(f) = \int_{M} \sigma(f)\,\Omega
\end{equation}
where $\sigma$ is a concave function with $\sigma(0)=0$, for instance $-f\ln f$. This entropy evolves, using the Liouville equation \eqref{eq.Liouville}, as
\begin{equation}
    \frac{\diff S}{\diff \lambda}
    = \int_{M} \sigma'(f)\frac{\partial f}{\partial \lambda}\,\Omega 
   = -\int_{M}\XX_H(\sigma(f))\,\Omega
     + 4\int_{M} f\sigma'(f)\,\frac{\partial H}{\partial\phi}\,\Omega.
\end{equation}
On a closed phase space, integrating $\XX_H(\sigma(f))$ by parts gives $\int_M \XX_H(\sigma(f))\,\Omega = -\int_M \sigma(f)\,\mathrm{div}(\XX_H)\,\Omega= 4\int_M \sigma(f)\,\frac{\partial H}{\partial\phi}\,\Omega$, so
\begin{equation}\label{eq.entropy.rate}
    \frac{\diff S}{\diff \lambda}
   = 4\int_{M}\bigl[f\sigma'(f)-\sigma(f)\bigr]\frac{\partial H}{\partial\phi}\,\Omega.
\end{equation}
Since $\sigma$ is concave with $\sigma(0)=0$, we have that $f\sigma'(f)-\sigma(f)\leq 0$,
and therefore the sign of entropy production is determined purely by the sign of
$\partial H/\partial\phi$. If $\partial H/\partial\phi$ is negative, then entropy grows, while if it is positive, entropy decreases.

\section{Applications}\label{sec.app}

In this Section, we show how the contact relativistic framework reduces to special relativistic motion of particles, to Newtonian gravity, and how it works when the particles are decaying and thus losing mass.

\subsection{Special Relativity}
Special relativity is recovered by choosing the flat Minkowski metric
\begin{equation}
    g_{\mu\nu} = \mathrm{diag}(-1,1,1,1), \qquad 
    g^{\mu\nu} = \mathrm{diag}(-1,1,1,1).
\end{equation}
With this choice, the mass--shell constraint \eqref{eq.constraint} becomes
    $-m^2 c^2 = -p_t^{\,2} + p_x^{\,2} + p_y^{\,2} + p_z^{\,2}$.

From Eq.~\eqref{eq.pu}, which relates momentum to four--velocity, we have
\begin{equation}
    p_t = -m \frac{\diff q^t}{\diff \tau}
        = -m c\,\frac{\diff t}{\diff \tau},
    \qquad
    p_x = m \frac{\diff q^x}{\diff \tau}
        = m \frac{\diff t}{\diff \tau} \frac{\diff q^x}{\diff t},
\end{equation}
and similarly for $p_y$ and $p_z$, where we used $q^t = ct$.
Introducing the three--velocity components
$v^i =  \diff x^i / \diff t$, the mass--shell condition reads
    $-m^2 c^2 = m^2 \gamma^2 (-c^2 + v^2)$ with $\gamma = \frac{\diff t}{\diff \tau}
    = \frac{1}{\sqrt{1 - v^2/c^2}}$.

The equations of motion in special relativity follow directly either from the
proper--time formulation \eqref{eq.evo.pq} or, equivalently, from the geodesic
equation in flat spacetime, $    \diff^{2} q^\mu / \diff \tau^{2} = 0 $.

\subsection{Newtonian Gravity}
We choose
a weak–field metric of the form (as in textbook \cite{hartle})
\begin{equation}\label{eq.metric.newton}
    g^{\mu\nu}
    = \mathrm{diag}\!\left(
        -1 + \frac{2\varphi}{c^{2}},
        \;1 - \frac{2\varphi}{c^{2}},
        \;1 - \frac{2\varphi}{c^{2}},
        \;1 - \frac{2\varphi}{c^{2}}
      \right),
\end{equation}
where $\varphi=\varphi(q)$ is the Newtonian gravitational potential.
This form of the metric corresponds to the standard weak–field, slow–motion
approximation in general relativity, where gravitational effects manifest
primarily through small perturbations in the metric components. Using the metric \eqref{eq.metric.newton}, and keeping only the leading
contributions in $\varphi/c^{2}$, the evolution equation for the momentum
reduces to
\begin{equation}
    \frac{\diff p_\mu}{\diff \tau}
        = - m\, \frac{\partial \varphi}{\partial q^\mu}.
\end{equation}
This is precisely the Newtonian equation of motion obtained from the
gravitational potential $\varphi$, showing that the relativistic contact
Hamiltonian framework reproduces Newton's gravitational law in the appropriate
limit.

In particular, for the spatial components ($\mu = 1,2,3$), the above equation
reduces to the familiar form
\begin{equation}
    m\,\frac{\diff^{2} q^i}{\diff t^{2}}
        = -\,\frac{\partial \varphi}{\partial q^{i}},
\end{equation}
once proper time is related to coordinate time in the slow–velocity limit.
Thus, Newtonian gravity emerges naturally from the contact formulation when the
metric is taken in its weak–field, non–relativistic approximation.

\subsection{Decaying Particles}
Suppose now that the particle is not just one elementary particle, but a blob of particles, such as a molecule or a macroscopic object. If the blob exchanges energy with its environment (for instance by emission or absorption), its effective rest mass varies along the worldline. We formulate this geometrically as a dependence of the rest mass on the proper time, $m(\tau)$.

The equations of motion \eqref{eq.evo.pq} for decaying particles are then
\begin{subequations}\label{eq.evo.pq.aging.tau}
\begin{eqnarray}
    \frac{\diff q^\mu}{\diff \tau}
        &=& \frac{g^{\mu\nu}p_\nu}{m(\tau)}, \\
    \label{eq.evo.p.aging.tau.p}\frac{\diff p_\mu}{\diff \tau}
        &=& -\frac{1}{2m(\tau)}\frac{\partial g^{\alpha\beta}}{\partial q^\mu}p_\alpha p_\beta
            +\frac{p_\mu}{m(\tau)}\frac{\diff m}{\diff \tau},
\end{eqnarray}
where we assumed that the metric be independent of $\phi$, as normally in general relativity, so that the only $\phi$-dependence in the Hamiltonian is through $m(\phi)$.
\end{subequations}

Let us now prescribe the usual exponential decay law for the mass,
\begin{equation}\label{eq.mass.exp}
    m(\tau)=m_0\,e^{-\alpha(\tau-\tau_0)},
    \qquad m_0>0,\ \alpha>0.
\end{equation}
Substituting into the momentum equation \eqref{eq.evo.p.aging.tau.p} then gives
\begin{equation}
    \frac{\diff p_\mu}{\diff \tau}
        = -\frac{1}{2m(\tau)}\frac{\partial g^{\alpha\beta}}{\partial q^\mu}p_\alpha p_\beta
          -\alpha p_\mu.
\end{equation}
Thus the decay rate rescales the four-momentum magnitude through a term parallel to $p_\mu$.

In the four-velocity formulation, the equations of motion are
\begin{subequations}\label{eq.evo.uv.aging.tau}
\begin{eqnarray}
    \frac{\diff q^\mu}{\diff \tau}
        &=& u^\mu, \\
    \frac{\diff u^\mu}{\diff \tau}
        &=& -\Gamma^\mu_{\alpha\beta} u^\alpha u^\beta,
\end{eqnarray}
\end{subequations}
where $\Gamma^\mu_{\alpha\beta}$ are the Christoffel symbols of the metric $g_{\mu\nu}$.
Indeed, with $p_\mu=m u_\mu$, the term $(\dot m/m)p_\mu$ in the momentum equation cancels exactly against $\dot m\,u_\mu$ from $\dot p_\mu=\dot m\,u_\mu+m\dot u_\mu$. Therefore, for metrics independent of $\phi$, mass decay does not introduce an additional drag in the geodesic equation: the worldline geometry is unchanged, while the momentum norm decreases because $m(\tau)$ decreases.

\paragraph{Kinetic theory for decaying particles.}
The single-particle description extends to an ensemble by introducing a distribution function $f(q,p,\phi)$ on the extended phase space $T^*Q\times\mathbb{R}$.  Applying the contact Liouville equation~\eqref{eq.Liouville} with the Hamiltonian~\eqref{eq.Ham}, the relevant $\phi$-derivative is
\begin{equation}\label{eq.dHdphi.aging}
    \frac{\partial H}{\partial \phi} = m(\phi)\,c^{2}\,m'(\phi).
\end{equation}
The contact Liouville equation for an ensemble of decaying particles therefore reads
\begin{equation}\label{eq.Liouville.aging}
    \frac{\partial f}{\partial \lambda}
    + \XX_H(f)
    - 4\,f m(\phi)\,c^{2}\,m'(\phi)
   =0.
\end{equation}
Even without collisions, the term $-4f\,\partial H/\partial\phi$ acts as an intrinsic source term. Particles in the distribution carry individual values of~$\phi$, thereby encoding a spread of lifetimes through the $\phi$-dependence of $m(\phi)$ as in~\eqref{eq.mass.exp}.

From the entropy-production formula~\eqref{eq.entropy.rate}, the sign of entropy change is governed by $m(\phi)\,m'(\phi)$.  For a particle that loses mass along its worldline---so that $m'(\phi)>0$, since $\phi$ decreases with proper time $\tau$ while the mass decreases with $\tau$---we have $\partial H/\partial\phi>0$, and the entropy of the distribution decreases, consistent with the energy being radiated away.  Conversely, a particle that absorbs energy has $m'(\phi)<0$, giving $\partial H/\partial\phi<0$ and entropy growth.

\subsection{Mass-less Particles}
If we started with Hamilton equations with derivatives with respect to the proper time on the left hand side, we would run into problems in the case of photons, since for mass-less particles the proper time is not defined \cite{landau2}. However, within the contact geometric framework, we can still describe the dynamics of mass-less particles consistently.

In the case of mass-less particles (setting $m=0$), such as photons, the contact Hamiltonian \eqref{eq.Ham} reduces to
\begin{equation}
    H(q,p,\phi)
    = \frac{1}{2}g^{\mu\nu}(q,\phi) p_\mu p_\nu.
\end{equation}
The mass shell condition $H=0$ gives $g^{\mu\nu}p_\mu p_\nu = 0$, the usual null condition.
The corresponding evolution contact equations are
\begin{subequations}\label{eq.con.can.H.massless}
\begin{eqnarray}
\frac{\diff q^\mu}{\diff \lambda} &=& 
    g^{\mu\nu} p_\nu, \\[0.3em]
\frac{\diff p_\mu}{\diff \lambda} &=&
    -\frac{1}{2}\frac{\partial g^{\alpha\beta}}{\partial q^\mu} p_\alpha p_\beta
    - \frac{p_\mu}{2}\frac{\partial g^{\alpha\beta}}{\partial \phi} p_\alpha p_\beta,
    \\[0.3em]
\frac{\diff \phi}{\diff \lambda} &=& 
g^{\mu\nu} p_\mu p_\nu.
\end{eqnarray}
\end{subequations}
For photons, we have $\diff\phi/\diff \lambda = g^{\mu\nu}p_\mu p_\nu = 0$ on the null shell, so $\phi$ is constant. If we are, moreover, in the standard setting where the metric does not depend on $\phi$, the parameter $\lambda$ serves directly as a parameter for the null geodesic, and the equations are
\begin{subequations}\label{eq.evo.pq.massless}
\begin{eqnarray}
    \frac{\diff q^\mu}{\diff \lambda}
        &=& g^{\mu\nu} p_\nu, \\[0.4em]
    \frac{\diff p_\mu}{\diff \lambda}
        &=& -\frac{1}{2}\frac{\partial g^{\alpha\beta}}{\partial q^\mu}
            p_\alpha
            p_\beta.
\end{eqnarray}
\end{subequations}
These equations describe null geodesics in curved spacetime without the need for reparametrization.

\paragraph{Kinetic theory for massless particles.}
An ensemble of photons is described by a distribution function $f(q,p,\phi)$ on the extended phase space.  For the massless Hamiltonian $H=\frac{1}{2}g^{\mu\nu}p_\mu p_\nu$, the $\phi$-derivative evaluates to
\begin{equation}
    \frac{\partial H}{\partial \phi}
    = \frac{1}{2}\frac{\partial g^{\alpha\beta}}{\partial \phi}\,p_\alpha p_\beta,
\end{equation}
and the contact Liouville equation~\eqref{eq.Liouville} specialises to
\begin{equation}\label{eq.Liouville.massless}
    \frac{\partial f}{\partial \lambda}
    + g^{\mu\nu} p_\nu\,\frac{\partial f}{\partial q^\mu}
    - \frac{1}{2}\!\left(\frac{\partial g^{\alpha\beta}}{\partial q^\mu}
      + p_\mu\,\frac{\partial g^{\alpha\beta}}{\partial \phi}\right)p_\alpha p_\beta\,
      \frac{\partial f}{\partial p_\mu}
    + g^{\alpha\beta}p_\alpha p_\beta\,\frac{\partial f}{\partial \phi}
    - 2f\,\frac{\partial g^{\alpha\beta}}{\partial \phi}\,p_\alpha p_\beta
    = 0.
\end{equation}
On the null shell $H=0$, the condition $g^{\alpha\beta}p_\alpha p_\beta=0$ kills the $\partial f/\partial\phi$ term, and $\phi$ is constant along each null geodesic (see~\eqref{eq.evo.pq.massless}).
For a metric that is independent of~$\phi$, all $\partial g^{\alpha\beta}/\partial\phi$ terms vanish, and \eqref{eq.Liouville.massless} reduces to
\begin{equation}
    \frac{\partial f}{\partial \lambda}
    + g^{\mu\nu} p_\nu\,\frac{\partial f}{\partial q^\mu}
    - \frac{1}{2}\frac{\partial g^{\alpha\beta}}{\partial q^\mu}\,p_\alpha p_\beta\,
      \frac{\partial f}{\partial p_\mu}
    = 0,
\end{equation}
which is the standard collisionless relativistic transport equation for photons in curved spacetime, i.e.\ the radiative transfer equation in the geometric-optics limit.  Since $\partial H/\partial\phi=0$ in this case, the entropy-production formula~\eqref{eq.entropy.rate} gives $\diff S/\diff \lambda=0$, in agreement with the reversibility of free photon propagation.

\section{Conclusion}
We have presented a contact-geometric formulation of relativistic particle dynamics in curved spacetime. By extending the traditional symplectic phase space to a contact manifold, we have incorporated the intrinsic proper time of the particle as an additional dimension in the phase space. This approach naturally leads to the recovery of standard general relativistic equations of motion while providing a deeper geometric understanding of the role of proper time.

While recovering the standard geodesic equations for particles with constant mass, the contact framework also gives a consistent description of particles with time-varying mass, such as decaying particles. Moreover, the motion of massless particles is obtained without the need for reparametrization. Finally, we derive a contact kinetic theory for ensembles of particles, which includes an intrinsic source term in the Liouville equation and a corresponding entropy production formula.

In the future, we would like to address relativistic fluid mechanics in the context of contact geometry, and to add also the field equations for the metric.

\section*{Acknowledgment}

MP was supported by Czech Science Foundation, project 23-05736S. MP is a member of the Nečas Centre for Mathematical Modeling. 
We acknowledge the use of GPT-5.3 Codex for cleaning up the text and for assistance with the latex typesetting.


\end{document}